\documentclass[10pt,letterpaper]{article}
\usepackage[cmex10]{amsmath}
\usepackage{opex3}
\usepackage{array}
\usepackage{mdwmath}
\usepackage{mdwtab}
\usepackage{subeqnarray}
\usepackage{cases}
\usepackage{cite}
\usepackage{float}
\usepackage{graphicx}
\usepackage{caption}
\usepackage{subcaption}
\usepackage{braket}
\usepackage{epstopdf}

\begin{document}

\title{Full-Vector Multi-Mode Matching Analysis of Whispering-Gallery Microcavities}

\author{Xuan Du$^1$, Serge Vincent$^1$, Mathieu Faucher$^2$, Mary-Jos$\acute{\mbox{e}}$e Picard$^2$, Simon Ayotte$^2$, Martin Guy$^2$, and Tao Lu$^{1,*}$}%

\address{1. Department of Electrical and Computer Engineering, University of Victoria, EOW 448, 3800 Finnerty Rd., Victoria, BC, V8P 5C2, Canada.\\
2. TeraXion, 2716 Einstein Street, Quebec, Quebec, G1P 4S8, Canada.}
\email{$^*$taolu@ece.uvic.ca}
\homepage{http://www.ece.uvic.ca/{\textasciitilde}taolu} %
\thispagestyle{empty}
\pagestyle{empty}
\begin{abstract}
We outline a full-vectorial three-dimensional multi-mode matching technique in a cylindrical coordinate system that addresses the mutual coupling among multiple modes copropagating in a perturbed whispering-gallery-mode microcavity.  In addition to its superior accuracy in respect to our previously implemented single-mode matching technique, this current technique is suitable for modelling waveguide-to-cavity coupling where the influence of multi-mode coupling is non-negligible. Using this modelling technique, a robust scheme for hybrid integration of a microcavity onto a silicon-on-insulator platform is proposed.
\end{abstract}


\bibliographystyle{osajnl}
\bibliography{IEEEfull,paper}

\section{Introduction}

Whispering-gallery-mode (WGM) microcavities display the uniquely combined features of ultrahigh quality factors (Q's) and small mode volumes. This has attracted research interest towards a wide range of applications, spanning single-molecule biosensing to on-chip optical filters~\cite{Armani2003,Vollmer2008,Dantham_Label,Lu12042011,Sun2009,Bahl2012,Lee:07,ilchenko2006,santiago2011nanoparticle}. Utility with regards to silicon photonics, however, is limited due to the relatively low quality factors of the microcavities available on silicon photonic platforms (e.g. silicon-on-insulator or SOI). Due to fabrication inaccuracies and the state of today's technology, the highest quality factor for a microcavity fabricated on a SOI wafer is reported to be $5{\times}10^6$\cite{Borselli2005a}. To implement a cavity with a quality factor above $10^6$, such as a silica microtoroid reported in\cite{Armani2003}, a hybrid integration method is required\cite{Hossein-Zadeh:07}. In the search for a optimal hybrid integration scheme with minimal alignment requirements, numerical modelling then becomes essential. Several techniques (with varied pros and cons) have been developed in the past to model cavity-to-waveguide coupling\cite{219420,toroid_Gorodetsky_Optical,Shirazi:13,Xu:13,Kaplan2013,Zou2011Quick} as well as to analyze general properties of microcavities\cite{toroid_Oxborrow_Traceable,Wiersig2003,PhysRevA.85.031805,Teraoka2003,Foreman2013a}. We have recently demonstrated a mode matching method (MMM) that models the above problems~\cite{Du2013a}. In this article, matching of single whispering-gallery modes propagating along the azimuthal direction of a cavity perturbed by nanoparticles was performed.  Such a technique yields highly accurate results when the dominant portion of light is confined within a single whispering-gallery mode on the course of propagation.  Here, we generalize the MMM analysis by incorporating multiple WGM's copropagating in the perturbed region of a cavity.  The validity of the newly developed algorithm is illustrated by investigating a tapered waveguide coupled to a silica microtoroid, wherein the mutual coupling between WGM's invalidates the single-mode matching method. This modelling technique allows for the examination of various schemes for hybrid integration of a whispering-gallery microcavity onto an SOI platform and for the identification of a robust integration approach.

\section{Theoretical formulations}
An ideal whispering-gallery microcavity (with refractive index independent of the azimuth $\phi$) may support multiple transverse modes of the same azimuthal order $M$. Consequently, continuous wave (CW) laser light delivered to the cavity excites a specific mode if its optical wavelength coincides with the resonance wavelength of that mode. The corresponding electric field distribution of an $l^{th}$ order transverse mode can be expressed as ${\bf E}({\bf r})=A{\hat {\tilde {\boldsymbol e}}}_{M,l}(\rho,z)e^{j{\tilde m}_l\phi}$. The term ${\tilde m}_l=M+j{\tilde m}_{i,l}$ is a complex number whose real part is an integer $M$ representing the azimuthal order of the WGM and whose imaginary part ${\tilde m}_{i,l}$ characterizes the intrinsic quality factor of the mode according to $Q=M/{2{\tilde m}_{i,l}}$. The unit vector ${\hat {\tilde e}}_{M,l}(\rho,z)$ is the normalized $\phi$-independent mode field distribution such that the squared amplitude $|A|^2$ represents the total energy of the light stored in the cavity (in units of Joules). For convenience, we have chosen a cylindrical coordinate system $(\rho,\phi,z)$ defined in~\cite{Du2013a}. By neglecting the loss in an ultrahigh-Q cavity, one may find a complete orthonormal set of modes where the othonormality relation between two modes of transverse order $l$ and $k$ satisfies~\cite{Haus2000}
\begin{equation}\label{M5_orthonormality}
\Braket{\boldsymbol{\hat{\tilde e}}_l(\rho,z)|\boldsymbol{\hat{{\tilde e}}}_k(\rho,z)}=\pi\int\epsilon (\rho,z) \boldsymbol{\hat{{\tilde e}}}_l^*{(\rho,z)}\cdot\boldsymbol{\hat{{\tilde e}}}_k{(\rho,z)} {\rho}d{\rho}dz=\delta_{lk}
\end{equation}
Here, $\delta_{lk}$ is the Kronecker delta and bra-ket notation is used to define the inner product in the functional space spanned by the WGM's as $\Braket{a|b}=\pi\int\epsilon a^*\cdot b \rho d\rho dz$ where $a^*$ is the complex conjugate of $a$. For convenience, we also drop the subscript $M$ by limiting our discussion to the same azimuthal order $M$. Also note that in the case of a large cavity, where the field is tightly focused into a small spot of mean radius $R$ and proper normalization is adopted, the orthonormality condition \eqref{M5_orthonormality} can be approximated by a more commonly used expression $\int \boldsymbol{\hat{{\tilde e}}}_l^*{(\rho,z,)}\cdot\boldsymbol{\hat{{\tilde e}}}_k{(\rho,z)} d{\rho}dz=\delta_{lk}$ \cite{Du2013a}.

A non-ideal microcavity has a refractive index profile that depends on the azimuthal angle.  To formulate the multi-mode matching technique, we assume light with wavelength $\lambda$ close to a non-ideal cavity resonance wavelength propagates in the cavity. The non-ideal microcavity can be subdivided into many sections along the azimuthal direction, each of which can be approximately considered as part of an ideal cavity that shares the same refractive index profile of that section. At a cross section at an arbitrary angle $\phi_0$, one may find a complete orthonormal set of $N(\phi_0)$ WGM's $\{{\hat {\tilde {\boldsymbol e}}}_0(\rho,z,\phi_0),{\hat {\tilde {\boldsymbol e}}}_1(\rho,z,\phi_0),\ldots,{\hat {\tilde {\boldsymbol e}}}_{N(\phi_0)-1}(\rho,z,\phi_0)\}$ corresponding to the approximated ideal microcavity.  Thus, any field pattern at $\phi_0$ can be expressed as a linear superposition of these modes:
\begin{equation}\label{M5_expansion_phi_0}
{\bf E}(\rho,z,\phi_0)=\sum_{l=0}^{N(\phi_0)-1} A_l{(\phi_0)} \ket{\boldsymbol{\hat{\tilde e}}_l{(\rho,z,\phi_0)}}
\end{equation}
 After propagating an infinitesimal azimuthal angle $\delta\phi_0$,  the field at $\phi_0+\delta\phi_0$ evolves to
\begin{equation}\label{M5_prop_delta_phi}
{\bf E}(\rho,z,\phi_0+\delta\phi_0)=\sum_{l=0}^{N(\phi_0)-1} A_l{(\phi_0)}e^{jm_l(\phi_0,\lambda)\delta\phi_0} \ket{\boldsymbol{\hat{\tilde e}}_l{(\rho,z,\phi_0)}}
\end{equation}
where $m_l$ is a complex number for the $l^{th}$ mode at the operating wavelength $\lambda$ that can be estimated according to~\cite{Du2013a}. The field ${\bf E}(\rho,z,\phi_0+\delta\phi_0)$ can be expanded onto the new set of orthonormal modes at $\phi_0+\delta\phi_0$:
\begin{equation}\label{M5_expansion_phi_0_delta_phi}
{\bf E}(\rho,z,\phi_0+\delta\phi_0)=\sum_{l=0}^{N(\phi_0+\delta\phi_0)-1} A_l{(\phi_0+\delta\phi_0)}\ket{\boldsymbol{\hat{\tilde e}}_l{(\rho,z,\phi_0+\delta\phi_0)}}
\end{equation}
By equating the right hand sides of both \eqref{M5_prop_delta_phi} and \eqref{M5_expansion_phi_0_delta_phi}, taking the inner product of both sides with $\bra{\boldsymbol{\hat{\tilde e}}_k(\rho,z,\phi_0+\delta\phi_0)}$, and utilizing the orthonormality condition (i.e. \eqref{M5_orthonormality}), we obtain
\begin{equation}\label{Mode_matrix}
{\vec A}(\phi_0+\delta\phi_0)={\widetilde C}(\phi_0)\cdot{\widetilde D}(\phi_0)\cdot{\vec A}(\phi_0)
\end{equation}
where we define ${\vec A}(\phi)\equiv \{A_1(\phi),A_2(\phi),\ldots, A_{N(\phi)}(\phi)\}^T$ as an $N(\phi)$-dimensional vector whose $l^{th}$ element represents the expansion coefficient of the $l^{th}$ mode and $N(\phi)$ as the number of the approximated ideal cavity WGM's available at $\phi$. $\widetilde{C}(\phi_0)$ and $\widetilde{D}(\phi_0)$ are $N(\phi_0+\delta\phi_0)\times N(\phi_0)$ and $N(\phi_0)\times N(\phi_0)$ matrices whose $(j,k)$ elements are
\begin{equation}\label{M5_D_matrix}
\begin{array}{rcl}
\{\widetilde{C}(\phi_0)\}_{jk}&=&\Braket{\boldsymbol{\hat{\tilde e}}_j{(\rho,z,\phi_0+\delta\phi_0)}|\boldsymbol{\hat{\tilde e}}_k{(\rho,z,\phi_0)}}\\
\{\widetilde{D}(\phi_0,\lambda)\}_{jk}&=&e^{jm_k(\phi_0,\lambda)\delta\phi_0}\delta_{jk}
\end{array}
\end{equation}
To obtain the resonance mode of a non-ideal cavity, one may divide it into $N$ slices at $\phi=\{\phi_0,\phi_1,\ldots, \phi_{N-1}\}$ and propagate the field one round trip by progressively applying Eq.~\eqref{M5_D_matrix}
\begin{equation}\label{M5_A_2pi}
{\vec A}(\phi_0+2\pi)=\prod_{l=0}^{N-1} [\widetilde{C}(\phi_l,\lambda)\cdot \widetilde{D}(\phi_l)]\cdot {\vec A}(\phi_0)
\end{equation}
Note that a mode profile requires that ${\vec A}(\phi_0+2\pi)=e^{j2\pi m_l}{\vec A}(\phi_0)$ where multiple values of $e^{j2\pi m_l}$ can be obtained by solving for the eigenvalues of the overall matrix in Eq.~\eqref{M5_A_2pi}, each of which correspond to a transverse mode of the cavity. Also note that resonance only occurs at a wavelength $\lambda$ such that the real part of $m_l$ is an integer. One may estimate the resonance wavelength $\lambda_l$ and the quality factor $Q_l$ of a specific mode according to $\lambda_l = \lambda \frac{ \text{Re}\{m_l\}}{M}$ and $Q_l = \frac{M}{2 \text{Im}\{m_1\}}$ following the same argument in \cite{Du2013a}. The corresponding modal field distribution at $\phi_0$ can be obtained from the eigenvector ${\vec A}_l(\phi_0)$ as
\begin{equation}
\ket{\boldsymbol{\hat{e}}_l(\rho,z,\phi_0)}=\sum_p A_{l,p}(\phi_0)\ket{\boldsymbol{\hat{\tilde e}}_p(\rho,z,\phi_0)}
\end{equation}
 The eigenvector above is normalized such that $|{\vec A}_l|^2=\sum_pA_{l,p}^*A_{l,p}=1$. In a non-ideal cavity, the modal field distribution is $\phi$-dependent. To reiterate, one may obtain the field distribution at any azimuthal angle by progressively applying \eqref{Mode_matrix} starting from $\phi_0$.

It is worth pointing out that, when a single whispering-gallery mode remains, ${\vec A}$ becomes a scalar and \eqref{M5_A_2pi} reverts to its simplified form as seen in the single-mode propagation equation (i.e. Eq. (11) of \cite{Du2013a}).

\section{Application and discussion}
The merits of the presented algorithm can be shown by optimizing the coupling of light from a tapered SOI waveguide to an ultrahigh-Q silica microtoroid.  The SOI waveguide under investigation has a width of 500 nm and is formed via an upper silicon layer as well as a lower SiO$_2$ layer of 220-nm and 2-$\mu$m respective thicknesses. The refractive indices of silica (1.44462) and silicon (3.48206) are taken from the literature~\cite{MALITSON1965,Handboodofoptics}. In our model, we employ a  microtoroid that has a major radius of 45 $\mu$m and a minor radius of 5 $\mu$m.

\begin{figure}[H]\label{fig1}
    \centering
    \begin{subfigure}[b]{0.48\textwidth}
        \centering
        \includegraphics[width=\textwidth]{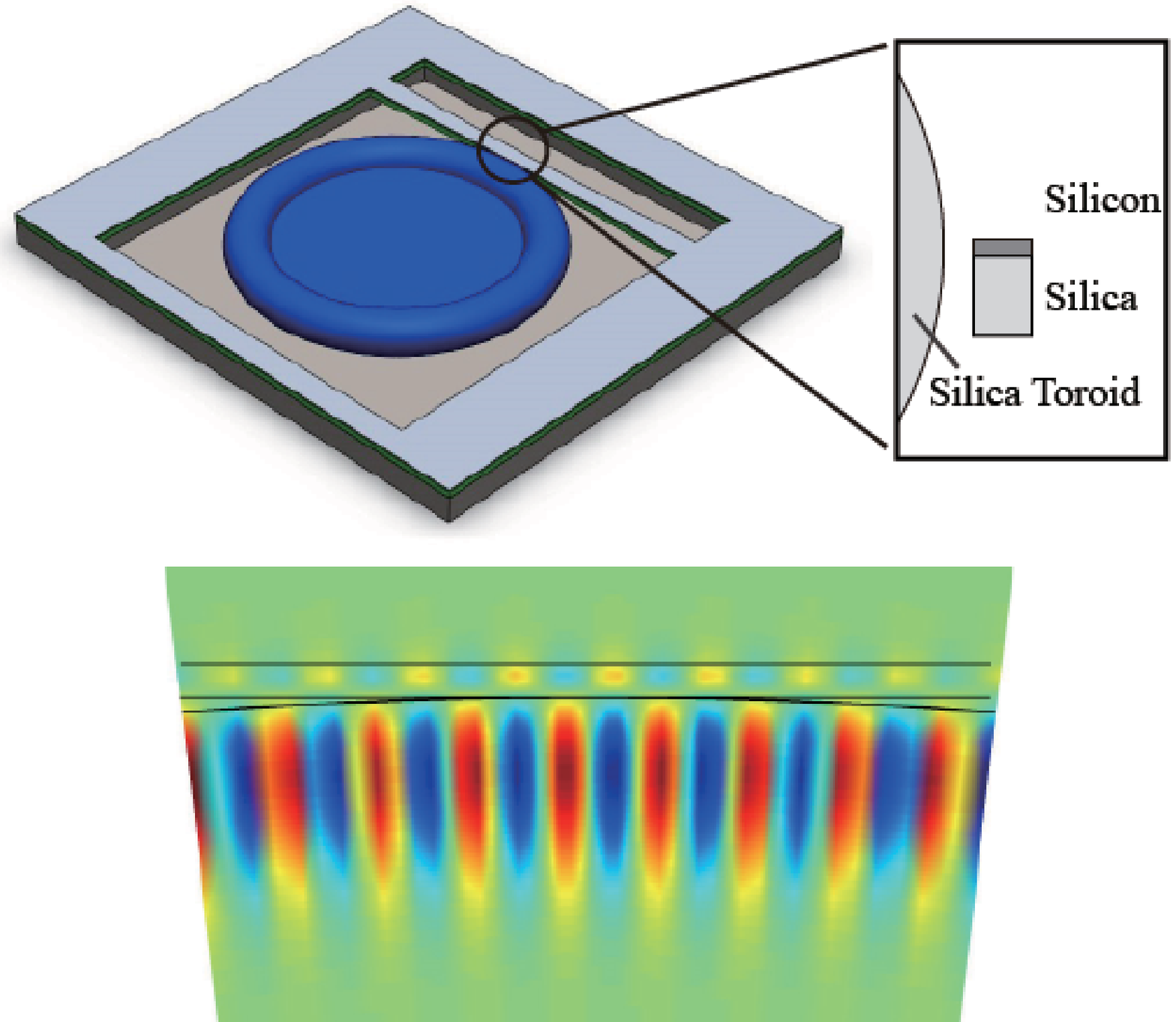}
        \caption{}
        \label{fig1_a}
    \end{subfigure}
    \begin{subfigure}[b]{0.48\textwidth}
        \centering
        \includegraphics[width=\textwidth]{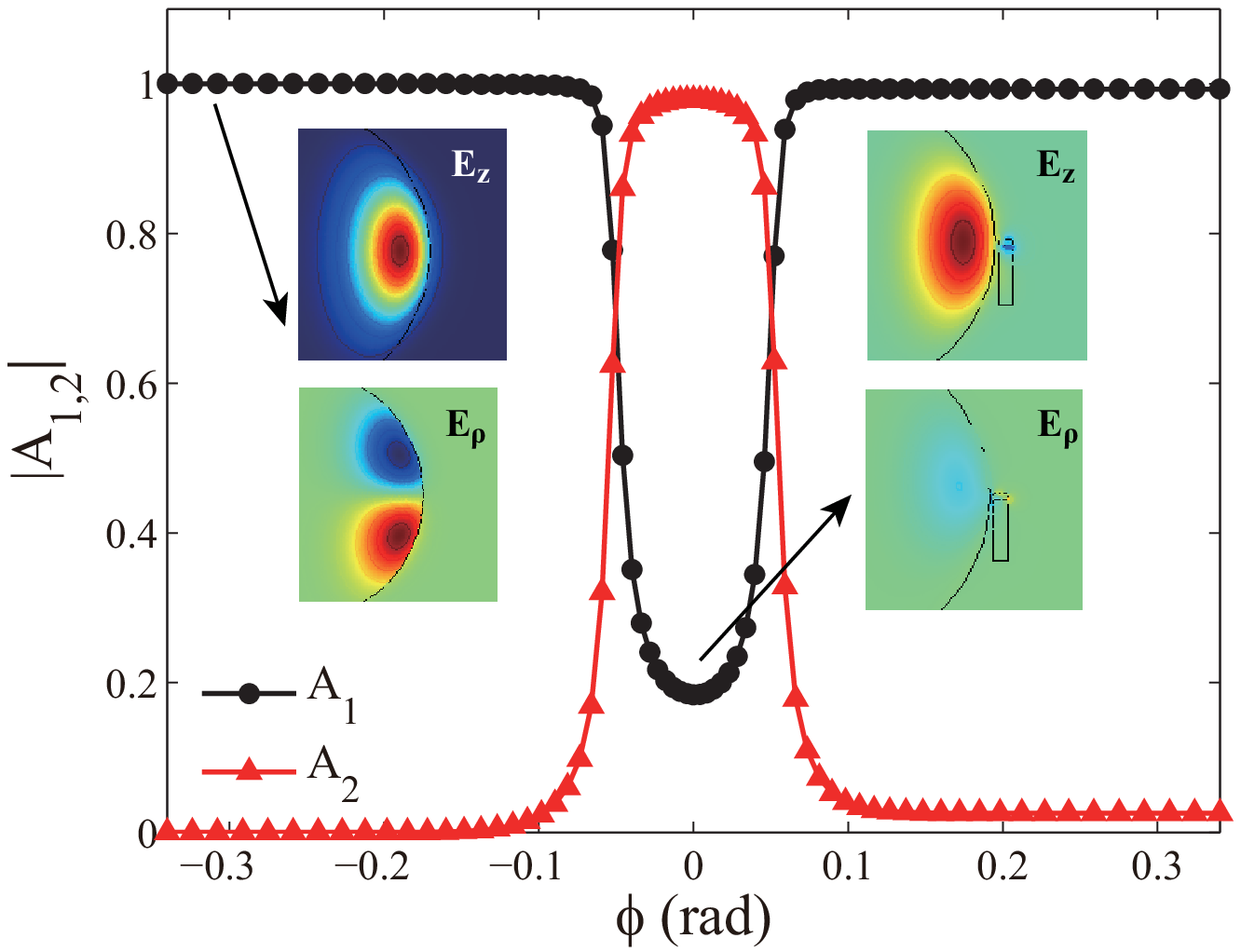}
        \caption{}\label{fig1_b}
    \end{subfigure}\\
    \begin{subfigure}[b]{0.48\textwidth}
        \centering
        \includegraphics[width=\textwidth]{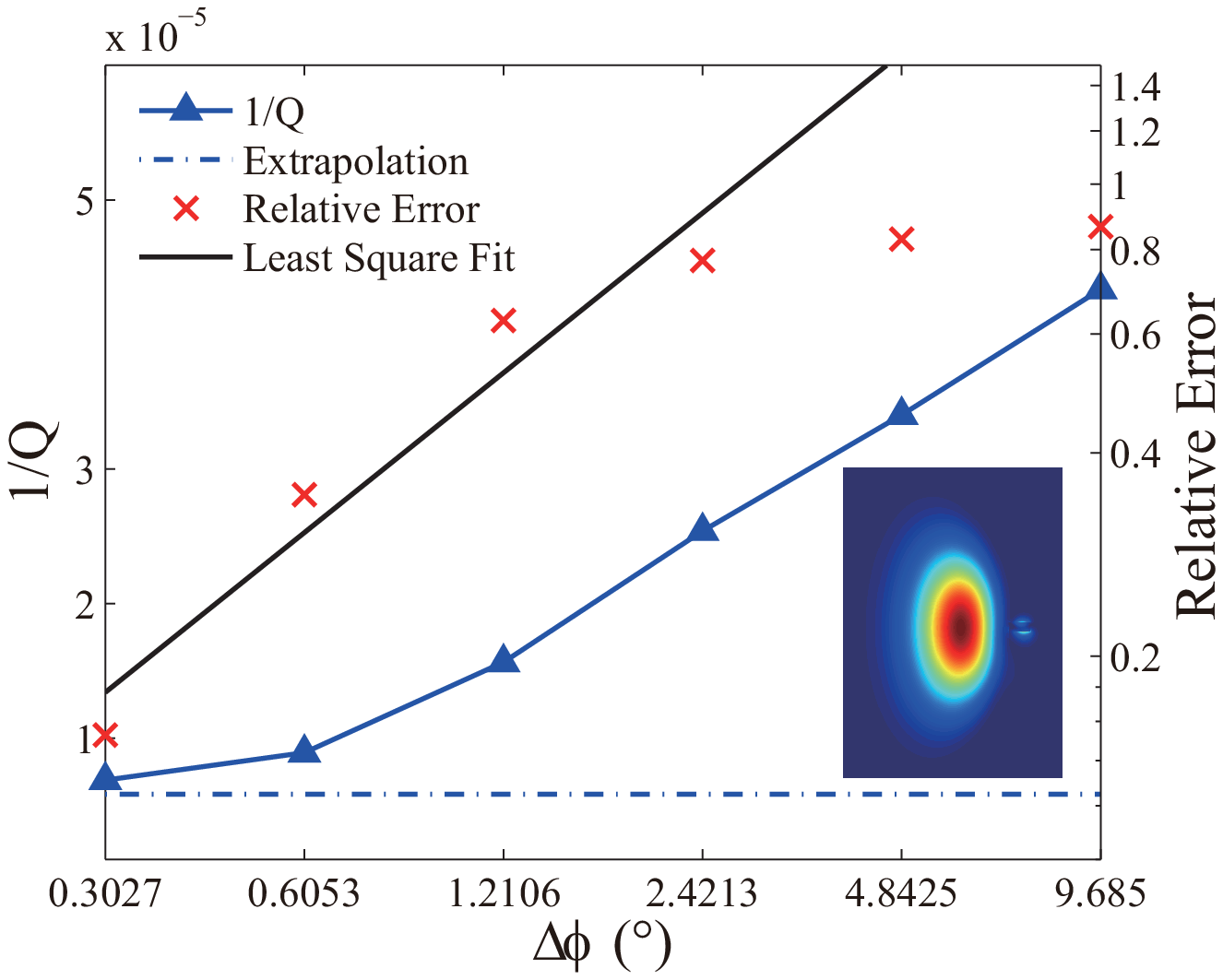}
        \caption{}\label{fig1_c}
    \end{subfigure}
    \begin{subfigure}[b]{0.48\textwidth}
        \centering
        \includegraphics[width=\textwidth]{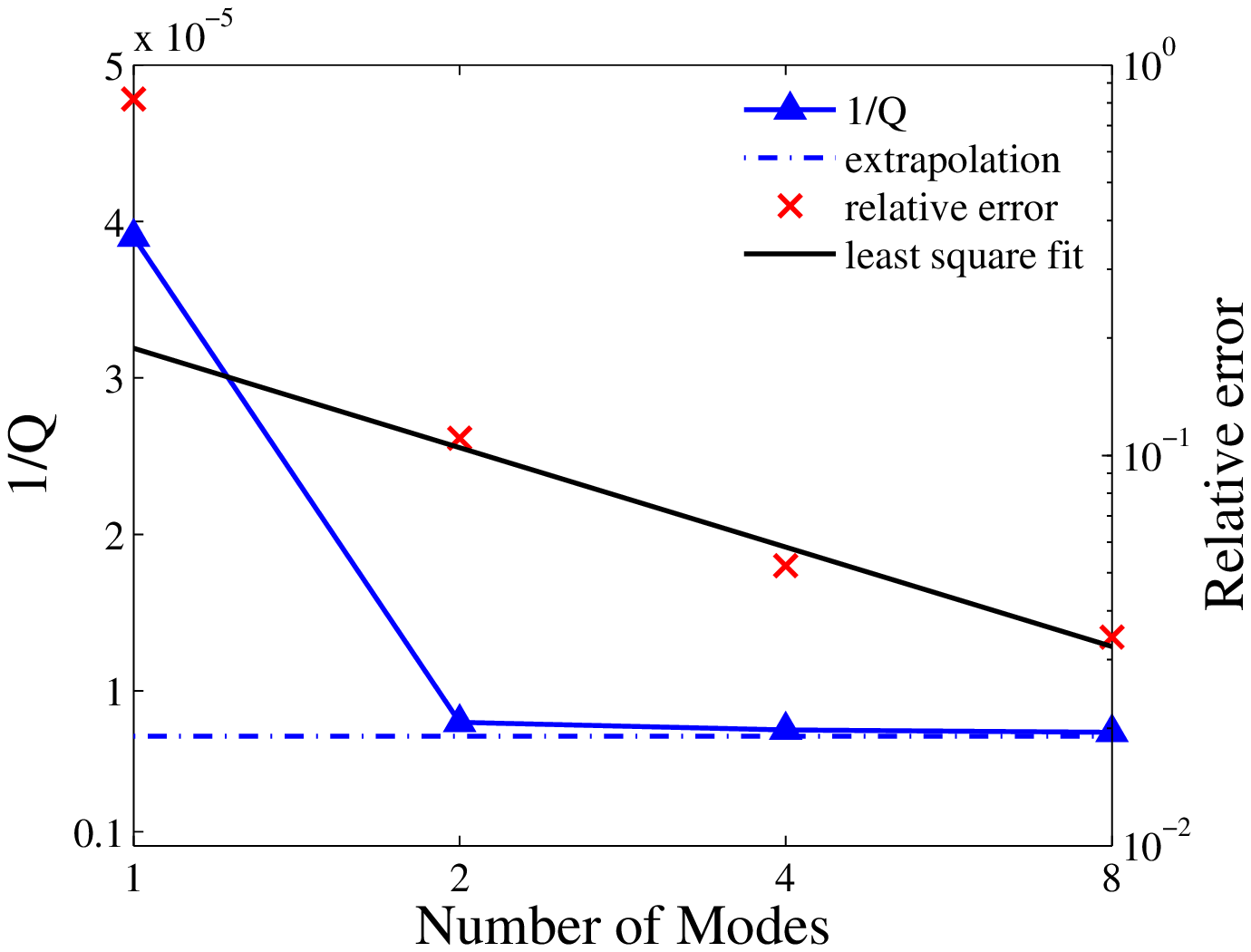}
        \caption{}\label{fig1_d}
    \end{subfigure}
        \caption{(a) The amplitude of the two lowest-order WGM's along the azimuthal direction when a straight SOI waveguide is placed at the equator of the cavity (i.e. left inset) and the top view of the major electric field distribution (i.e. right inset). (b) The reciprocal of the quality factor as a function of azimuthal angle step size, revealing a convergence rate of $O(\delta \phi^{0.78})$. The inset displays a intensity distribution at a cross section perpendicular to the azimuthal direction. (c) The reciprocal of the quality factor as a function of the number of modes included in the simulation, revealing a convergence rate of $O(\delta \phi^{0.8})$.}
\end{figure}

In this test case we study the coupling scheme between a straight waveguide and a toroid, as is shown in the left inset of Fig.~\ref{fig1_a}. Light propagation along the azimuthal direction is computed at a wavelength of 1500 nm. In the first simulation, we place the straight SOI waveguide in contact with the microtoroid's equator and a 289$^{th}$ fundamental quasi-TE toroid mode is launched at $\phi=-0.34$ rad, wherein the waveguide is 3 wavelengths away from the cavity. Beyond that point, the coupling between the cavity and the waveguide falls to a negligible level (with Q degradation below 10$^{-8}\%$) and so it is disregarded. The main plot in Fig.~\ref{fig1_a} shows the amplitudes of the first two lowest-order modes (quasi-TE and quasi-TM) along the azimuthal direction. In the intermediate region (around $\phi=-0.06$ rad to $\phi=0.06$ rad), these two modes become hybrid WGM-SOI modes and strong coupling takes place between them. The right inset of Fig.~\ref{fig1_a} further depicts a top view of the cavity-waveguide coupling while the right inset of Fig.~\ref{fig1_b} displays the cross-sectional intensity distribution that is perpendicular to the azimuthal direction. Clearly, energy is coupled back and forth between the first and second mode in a similar manner as that of a directional coupler. The field at the output end is carried by both modes with A$_1$=0.9928 and A$_2$=0.0256 (i.e. energy carried by higher-order modes is further diminished), thus resulting in a coupling Q-factor of $10^5$.

To characterize the convergence rate of our algorithm, we plot the reciprocal of the quality factor as a function of azimuthal angle steps and number of modes included (illustrated by the blue triangles in Fig.~\ref{fig1_b} and Fig.~\ref{fig1_c}). The expectation value of this quantity at an infinitesimal azimuthal step or infinite mode numbers (i.e. blue dashed line) was extracted through the Richardson extrapolation procedure\cite{Press1992} and the relative error (i.e. red cross markers) was estimated with the extrapolated value as a reference. As shown by the black lines in Fig.~\ref{fig1_b} and Fig.~\ref{fig1_c}, the inverse of the Q factor has a converge rate of $O(\delta \phi^{0.78})$ and $O(\delta \phi^{0.8})$ respectively.
To reach the critical coupling condition, wherein the coupling Q factor is equal to the intrinsic Q factor of the cavity (assumed to be a practical value of $2{\times}10^8$, as specified by the dot-dashed line of Fig.~\ref{fig1_b}), we gradually displaced the tapered waveguide horizontally away from the toroid. As displayed in Fig.~\ref{fig1_b}, a $\mbox{Q}_{\mbox{coupling}}$ of 10$^5$ is computed when the waveguide touches the cavity surface and 10$^{10}$ is computed at a gap size of 2.5 wavelengths. For a waveguide sitting in the equatorial plane, a gap size of 0.75 $\mu$m is also determined to be desirable in establishing critical coupling. It is evident from the figure that the coupling Q factor is sensitive to the gap distance. More specifically, a gap distance fluctuation on the order of a few hundred nanometers may cause an order of magnitude drop in the coupling Q factor. Therefore, to integrate a toroid onto a SOI platform with this scheme, impractically precise alignment is required.

\begin{figure}[H]\label{fig2}
    \centering
    \begin{subfigure}[b]{0.48\textwidth}
        \centering
        \includegraphics[width=\textwidth]{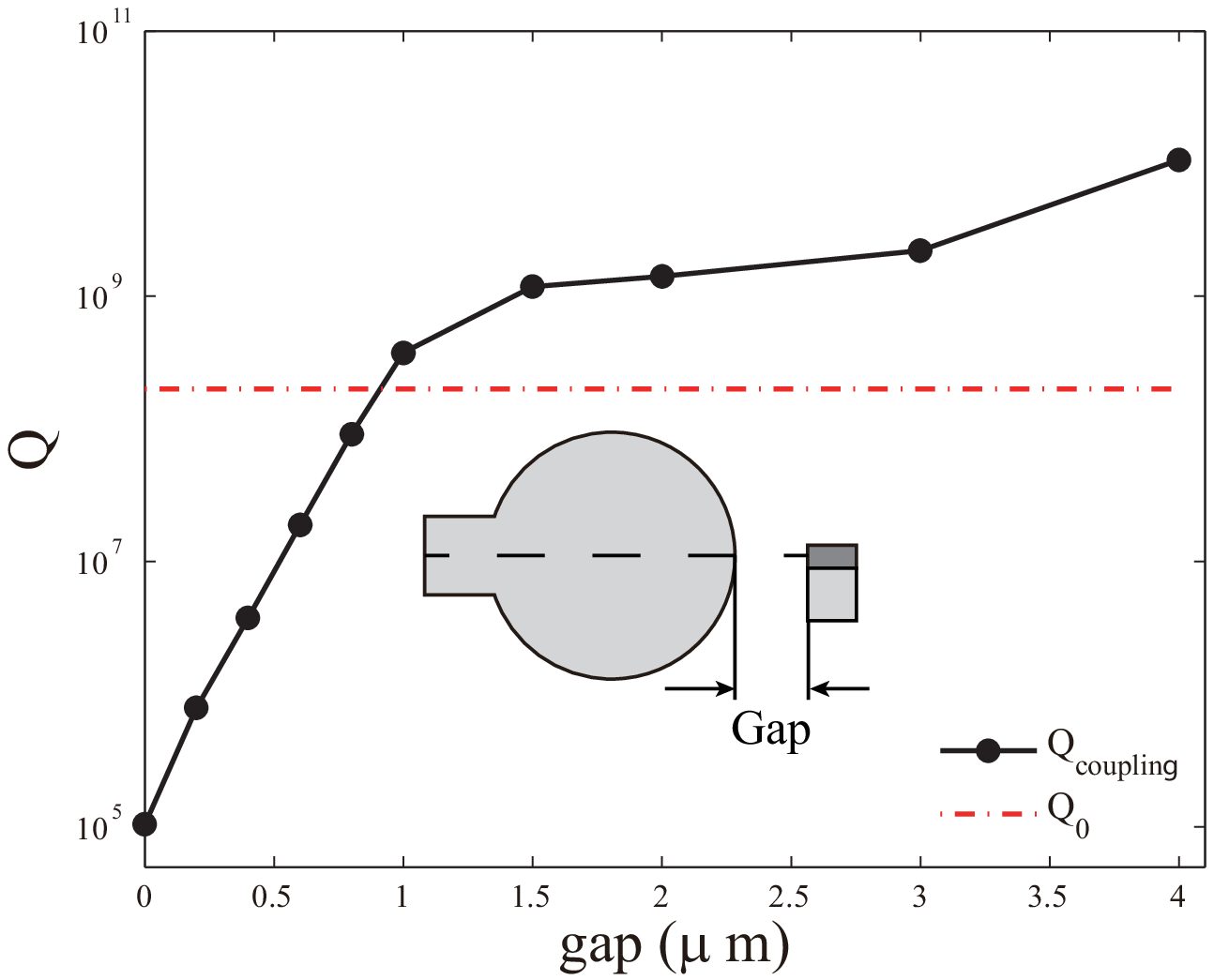}
        \caption{}\label{fig2_a}
    \end{subfigure}
    \begin{subfigure}[b]{0.48\textwidth}
        \centering
        \includegraphics[width=\textwidth]{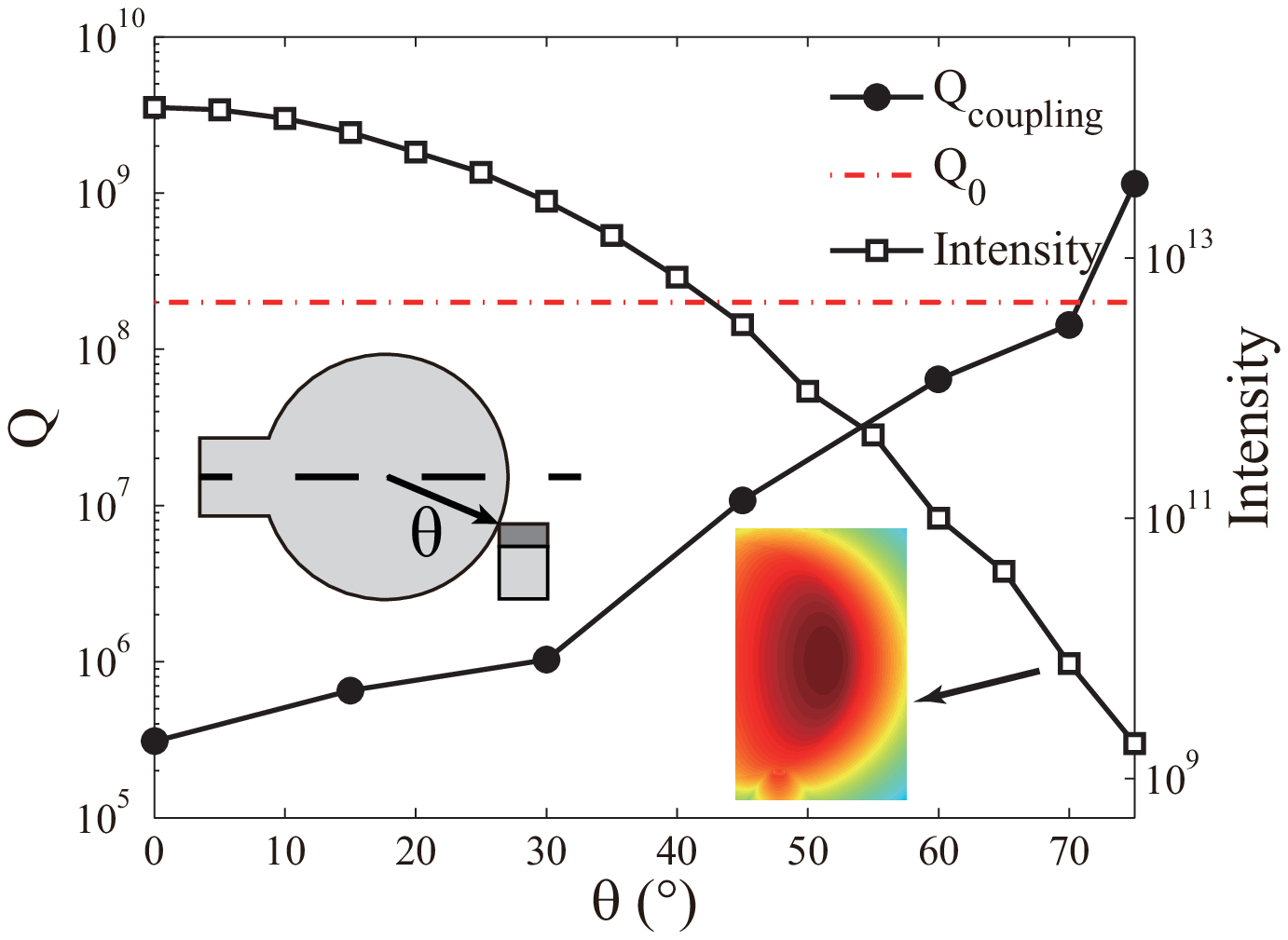}
        \caption{}\label{fig2_b}
    \end{subfigure}
    \caption{Coupling Q factor as a function of (a) gap distance and (b) vertical angle.}
\end{figure}

To circumvent the alignment restriction, one can keep the waveguide in contact with the cavity and move it vertically by adjusting the thickness of the insulator layer as to tune the value of the coupling Q factor. In such a configuration, one may pinpoint the location of the waveguide via an angle $\theta$ in respect to the equatorial plane, as illustrated in the inset of Fig.~\ref{fig2_b}. The Q factor as a function of $\theta$ is shown in the main plot as solid circle markers, wherein a Q-factor of $3\times10^5$ is calculated at $\theta=0$. It is also worth mentioning that the $\theta=0$ position (corresponding to the top surface of the SOI waveguide being aligned with the equator) is different from that in the previous case (i.e. the midplane of the silicon layer being aligned with the equator). At a larger $\theta$, the local field intensity of the cavity mode is smaller. As a result, coupling between the cavity and the waveguide is weaker and a larger coupling Q factor is observed. The right axis of Fig.~\ref{fig2_b} represents the field intensity on the surface of the cavity at different $\theta$. It is important to note that the Q factor function inversely resembles the intensity curve. For a straight SOI waveguide physically touching the toroid surface and that is placed below the equator, a waveguide location at $\theta=65^\circ$ is favourable in achieving critical coupling.
\section{Conclusion}
A three-dimensional full-vector multi-mode matching method was formulated in cylindrical coordinates and tested. This technique solves generalized whispering-gallery-mode cavity problems, yet in this paper was specifically applied to a toroid-SOI coupling geometry where no precise control of the gap distance between the cavity and the waveguide was required. Experimental verification of this integration scheme is forthcoming. Once again, the multi-mode matching method proposed in this article is applicable to numerous classes of WGM related scenarios where the energy transfer between different modes is non-negligible.

\end{document}